# Active Temporal Multiplexing of Indistinguishable Heralded Single Photons

C. Xiong[1*], X. Zhang[1], Z. Liu[2,3], M. J. Collins[1], A. Mahendra[1,2], L. G. Helt[4], M. J. Steel[4], D.-Y. Choi[5], C. J. Chae[6], P. H. W. Leong[2], and B. J. Eggleton[1].

[1]Centre for Ultrahigh bandwidth Devices for Optical Systems (CUDOS), Institute of Photonics and Optical Science (IPOS), School of Physics, University of Sydney, NSW 2006, Australia.

[2]School of Electrical and Information Engineering, University of Sydney, NSW 2006, Australia.

[3]Department of Communication Engineering, School of Information Engineering, Guangdong University of Technology, China.

[4]CUDOS and MQ Photonics Research Centre, Department of Physics and Astronomy, Macquarie University, NSW 2109, Australia.

[5]CUDOS, Laser Physics Centre, Australian National University, Canberra, ACT 2601, Australia

[6]NICTA-VRL, University of Melbourne, VIC, Australia (now with Advanced Photonics Research Institute, Korea)

[*]Correspondence to: chunle.xiong@sydney.edu.au.

**It is a fundamental challenge in quantum optics to deterministically generate indistinguishable single photons through non-deterministic nonlinear optical processes, due to the intrinsic coupling of single- and multi-photon generation probabilities in these processes. Actively multiplexing photons generated in many temporal modes can decouple these probabilities, but key issues are to minimize resource requirements to allow scalability, and to ensure indistinguishability of the generated photons. We demonstrate the multiplexing of photons from four temporal modes solely using fiber-integrated optics and off-the-shelf electronic circuits. We show a 100% enhancement to the single photon output probability without introducing additional multi-photon noise. Photon indistinguishability is confirmed by a four-fold Hong-Ou-Mandel quantum interference with a 91% visibility. Our demonstration paves the way for scalable multiplexing of many non-deterministic photon sources to a single near-deterministic source, which will be of benefit to future quantum photonic technologies.**

Single particles of light — photons — are a vital resource for the implementation of quantum-enhanced technologies such as optical quantum computing [1] and simulation [2]. To make such technologies practical requires ideal single photon sources, which can emit single photons on-demand, and indistinguishable in all relevant degrees of freedom—central frequency, bandwidth, spatial mode and polarization [3, 4]. Despite recent progress on relaxing these requirements [5, 6], sources that meet the required thresholds do not yet exist. Two strategies have been proposed to develop the desired photon sources [7]. One is to use "single-emitter" quantum systems [8-11] such as quantum dots or color centers in diamond. These systems typically emit single photons nearly on-demand [12], but producing highly indistinguishable photons from distinct emitters remains challenging due to the difficulty of fabricating identical emitters at nanoscale [13,14]. The alternative approach is to generate correlated photon pairs via spontaneous nonlinear optical processes, such as parametric down conversion or four-wave mixing in suitable crystals or waveguides, where the detection of one photon in a pair "heralds" the existence of its partner [15-17]. However the photon pair generation events are unpredictable (being associated with vacuum fluctuations) and contain contributions from multi-pair events. Indeed the probabilities of single- ($P_1$) and multi-pair ($P_{>1}$) events are both related to the mean number of pairs created per pump pulse $\mu$. They both increase with $\mu$, and $P_{>1}$ increases more rapidly (to leading order it grows quadratically rather than linearly). Therefore these sources are usually operated in the $\mu \ll 1$ (and thus $P_1 \ll 1$) regime to minimize the multi-photon noise. Unfortunately, most useful quantum protocols require many simultaneous single photon inputs in different modes, and as the success rate falls as $(P_1)^n$ for $n$ input modes, operation quickly becomes impractical [4, 5]. This has limited the world record for quantum photonic demonstration to the eight-photon level [18].

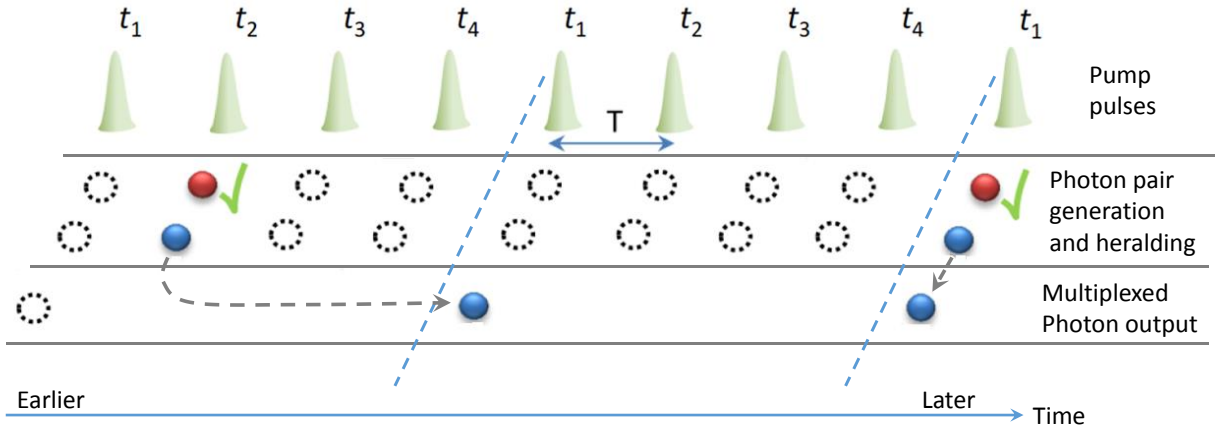

**Fig. 1.** The principle of active temporal multiplexing. A nonlinear device is pumped by pulses separated in time by period $T$, each generating correlated photon pairs randomly. The two photos from each pair are spatially separated by frequency (color) and the heralding photons (red) are detected, indicating the existence of the heralded photons (blue). Depending on in which time bin a pair is generated, an appropriate delay is applied to the heralded photon so that it always appears in time bin $t_1$ with a nominal period $NT$ ($N=4$ in this work).

A promising solution is to actively multiplex non-deterministic photons in different spatial or temporal modes to enhance the probability of single-photon output [19-22]. Spatial multiplexing has been implemented in a few architectures [23-25] but scaling quickly becomes infeasible as the number of photon sources and heralding detectors increases rapidly with the number of modes to be multiplexed [25, 26]. Temporal multiplexing, proposed in [20-22], reuses the same detectors and photon generation components, and thus is significantly more resource efficient and scalable. The scheme in [22] requires an electronic circuit to extract timing information from the heralding photons, which is subsequently used to control a switching network that actively routes the heralded photons into a pre-defined temporal mode (Fig. 1). The challenges associated with this technique include: (I) precisely synchronizing the photons' path lengths, (II) managing their arrival time to the accuracy of several picoseconds, and (III) controlling their polarization to maintain the photons' indistinguishability; and (IV) the need for ultra-low-loss optical components so that the desired enhancement can be achieved. The scope of this work is to experimentally overcome all of these challenges using all-fiber integrated low-loss optical components and off-the-shelf fast electronic circuits, and to reveal the potential of this scheme for deterministic indistinguishable single photon generation. We aim to show a substantial increase in the heralded single photon output probability at a given clock cycle with no concomitant increase in the multi-pair contamination.

The principle of our demonstration is illustrated in Fig. 1. Compared to pump pulses at period $4T$, pump pulses at period $T$ are approximately four times as likely to generate a pair in the given time frame of $4T$ if the individual pulse energy is the same. However, the random nature of the generation process within each time bin remains the same. The situation changes after the heralded photons are actively delayed to time bin $t_1$: if the switching network has sufficiently low losses, the heralded single photon output probability at the $4T$ clock period will be increased.

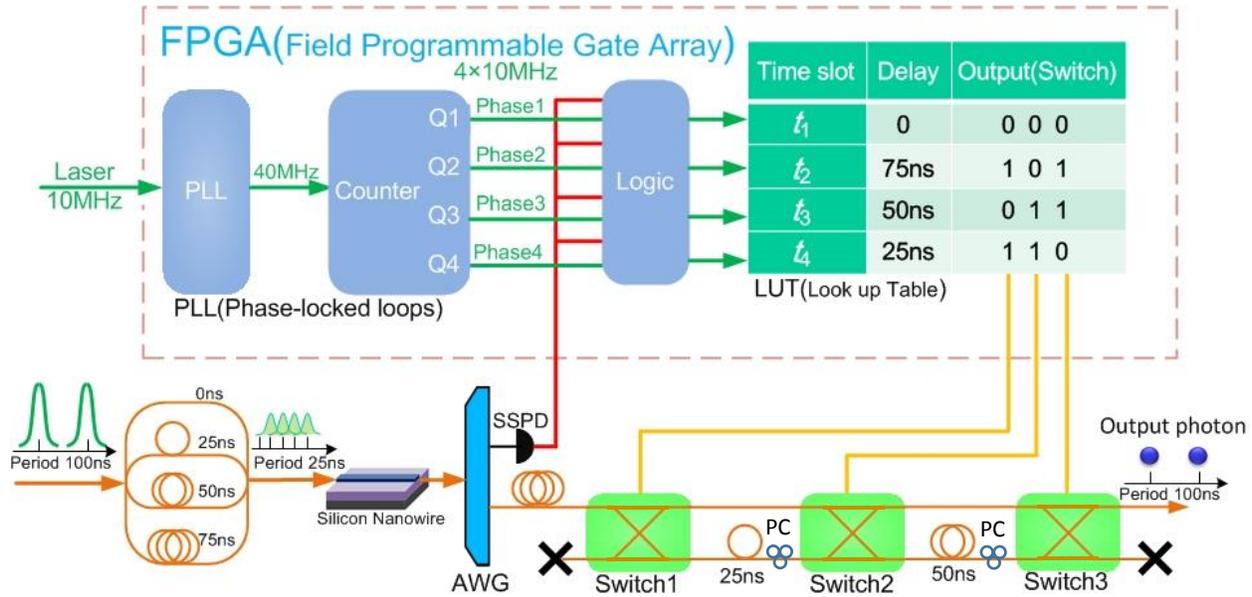

**Fig. 2.** Experimental setup of four temporal mode multiplexing. Pulses from a mode-locked picosecond fiber laser were split to four copies using fiber couplers and tunable delay lines, and pumped a silicon nanowire for spontaneous four-wave mixing. The 0, 25, 50 and 75 ns delays are all relative to the uppermost optical path. After pump blocking, frequency selection, and spatial separation of the two photons of each pair, the heralding signals were analyzed by a FPGA and the heralded photons were buffered using a long fiber delay to wait for the electronic decisions. The loss of the 200 m long buffer fiber was less than 0.1 dB. The FPGA configures the switching network to route the heralded photons into a single spatial-temporal mode. Logic '0' means the photon remains in the input ("bar") channel; a '1' means the photon is routed to the cross channel.

To implement the scheme shown in Fig. 1, we designed an experiment as shown in Fig. 2 (see Fig. S1 for the full setup). A mode-locked fiber laser with a repetition rate of 10 MHz (100 ns period) produces 10 ps pulses at 1550 nm. Each pulse was split into four pulses spaced by 25 ns using two 1-to-4 fiber couplers and three tunable optical fiber delay lines. The four pulses then propagated along a 3 mm long nonlinear silicon nanowire, probabilistically generating correlated photon pairs via spontaneous four-wave mixing in the four time bins [27]. Due to energy conservation and phase matching, photon pairs were generated at frequencies symmetrically around the pump over a 6 THz bandwidth [28]. An arrayed waveguide grating (AWG, 100 GHz channel spacing and 50 GHz channel bandwidth) was used to select the photon pairs generated at 1545 and 1555 nm, block the pump, and spatially separate the two photons of each pair. The 1555 nm photons were detected by a fast and low-noise niobium nitride superconducting single photon detector (SSPD) as heralding signals. These signals contained the timing information of the 1545 nm photons and were sent to a field-programmable gate array (FPGA) for analysis.

A phase-locked loop in the FPGA was used to lock to and multiply the laser's original 10 MHz clock to a 40 MHz clock. A finite state machine operating on the 40 MHz clock generated 4 non-overlapping clocks at 4 phases relative to the 10 MHz clock. A heralding photon detection signal from the SSPD was ANDed with each clock phase and an appropriate three binary-digit output latched. The output was connected to the switching network so that the 1545 nm photons were routed into the appropriate sequence of delay lines. All of these operations require the careful alignment of the clock with the optical pulses that contain the generated photons. This was done by optimizing the counts in a series of coincidence measurements, adjusting the tunable optical delay lines and tunable digital delays [28].

To receive any benefit from a four temporal mode multiplexing setup, the switching network must have a total loss below the four times (i.e. 6 dB) maximum expected enhancement. We used optical ceramic switches, made from ultra-low loss lead lanthanum zirconium titanate [25]. These switches were fiber pigtailed and spliced with the fiber delay lines to minimize the loss of each path to 2.8 dB on average. The setup described so far ensures indistinguishability in the spectral and temporal degrees of freedom [27], but we also require indistinguishability in polarization. In Fig. 2, the heralded photons from different time bins have the same polarization before they enter the switching network. However they experience different optical paths to obtain the correct delays, and to minimize losses these components are not polarization maintaining. This was addressed using two polarization controllers (PCs) applied to the two optical delay lines [27]. The additional loss introduced by each PC was about 0.1 dB.

The key to verifying our design is to compare the heralded single photon output probability per 100 ns clock period (i.e. the original 10 MHz clock) at the same multi-photon noise level for sources with and without the multiplexing switching network. These two quantities were characterized by coincidence-to-accidental ratio (CAR) measurements [29]. When a pair of photons generated in the same pump pulse are detected and the detection signals sent to a time interval analyzer (TIA), a coincidence is recorded. When photons generated from different pulses are detected, the coincidence represents an accidental coincidence. All of these coincidences ($C$) and accidentals ($A$) are recorded as a histogram by the TIA [27, 29], and CAR=$C/A$. The measured CAR as a function of the coincidence rate without multiplexing (NO MUX, i.e. pumping at 10 MHz) is plotted in Fig. 3A, indicated by diamonds. The CAR decreases with the increased coincidence rate due to multi-pair noise and this is a typical feature of such measurements [25, 28, 29]. For comparison, we performed measurements at the same pump peak powers for the multiplexed source (MUX, i.e. pumping at 40 MHz and adding the switching network to the setup). The results are plotted as triangles in Fig. 3A. The CAR still decreases with the increased coincidence rate because the original NO MUX sources have this feature. However when compared with the NO MUX source, at the same CAR, i.e. the same multi-pair noise level, the coincidence rates are nearly doubled. As simply doubling the pump laser repetition rate can lead to similar results in Fig. 3A [28], we choose to express the improvement as an enhancement factor of MUX/NOMUX heralded single photon output probability per 100 ns clock period at the same CAR level. Taking into account the losses of waveguide-fiber coupling, spectral filters and the efficiency of detectors, we estimate the mean number of pairs per 100 ns clock period from the measured coincidence rate at each CAR level, and then infer the heralded single photon output probabilities using a thermal distribution function for both NO MUX and MUX sources [27]. The enhancement factor at each CAR level is plotted in Fig. 3B as circles, showing that our four temporal-mode multiplexing nearly enhances the heralded single photon output probability by 100% (i.e. 3 dB). Using the single count rate from the heralding detector, we infer the ideal enhancement to be approximately 4 times (i.e. 6 dB). The practical enhancement is less than the ideal due to the 3 dB loss of the switching network.

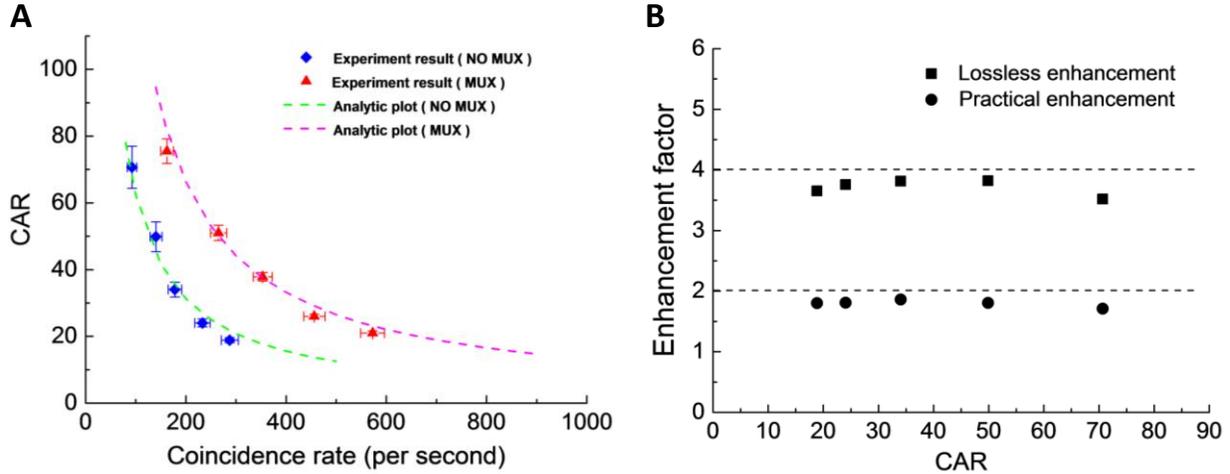

**Fig. 3.** Comparison between sources with and without multiplexing. (**A**) CAR as a function of coincidence rates. Poisson error bars are used for the plots. Dashed lines are analytic plots using the model in [25]. (**B**) The inferred ideal lossless and practical enhancement factors to the heralded single photon output probability at each CAR level.

For the multiplexed source to be useful, the multiplexed heralded photons must be indistinguishable. This was tested by Hong-Ou-Mandel (HOM) quantum interference [30]. We built another heralded single photon source based on a second 3 mm long silicon nanowire pumped by the same 10 MHz laser, but without multiplexing (Fig. S1). The photons from this second source were guaranteed to be in a defined spatial-temporal state, i.e., in an identical polarization state and at the accurate 100 ns clock cycle of the laser, and so they provided a reference to check if the multiplexed photons were indistinguishable. We firstly performed a two-fold interference measurement to find the appropriate delay between photons from the two sources [31]. In this measurement, the pump powers were set at a level of CAR=18 for both. The two-fold dip shows a raw visibility of 24% (diamonds in Fig. 4A). Then we took the four-fold HOM interference measurement, but at higher pump powers for both sources in order to have sufficient coincidence counts while maintaining a reasonable measurement time with our low-efficiency detectors. We observed a four-fold HOM dip with raw visibility of 69% (squares in Fig. 4A), indicating that non-classical interference occurred between the multiplexed photons and the photons from the second source. To check that the residual photon distinguishability is not because of multiplexing but due to multi-photon noise at high pump powers, we measured the detector dark count and multi-photon contribution from each source [27, 32]. Using these data we corrected the raw data, which yielded a visibility of 91% (Fig. 4B), clearly showing that the multiplexed photons are highly indistinguishable. The error bars are large mainly due to the low count rates resulting from the low detection efficiency.

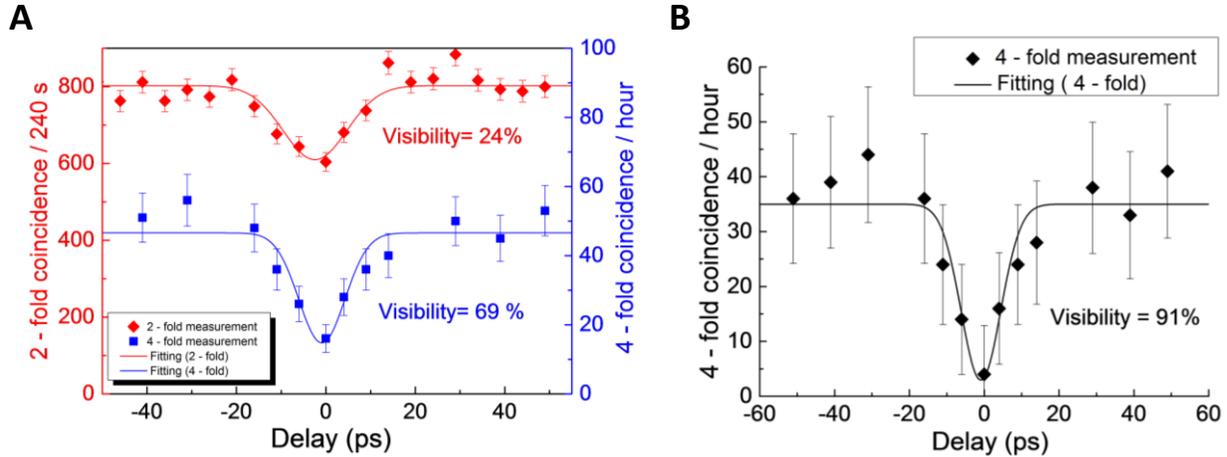

**Fig. 4.** Indistinguishability measurement of the multiplexed photons. (**A**) Raw visibility of two-fold (diamonds, left axis) and four-fold (squares, right axis) measurements. (**B**) Four-fold HOM dip visibility after subtracting multi-photon noise. Poisson error bars are used for the plots. Solid lines are Gaussian fits according to the spectral filtering shape in the experiment.

Bringing non-deterministic nonlinear photon sources into the deterministic regime is a multi-faceted task, mainly involving breaking the intrinsic statistical limit, improving spectral filtering losses and improving single photon detector efficiencies. We have demonstrated a resource efficient and thus scalable scheme that can largely decouple the probabilities of single and multi-photons and therefore overcome a fundamental statistical challenge. Every additional switch added to the switching network shown in Fig. 2 will double the number of temporal modes to be multiplexed, providing approximately 2 dB additional gain to the heralded single photon output probability (ideal 3 dB gain subtracting the 1 dB loss from the switch). Using low-loss optical fiber integrated components in the switching network makes it possible to scale this up. Together with the development of low-loss filters and high-efficiency detectors, this will ultimately provide a solution for photon sources required for optical quantum computing and simulation.

**References**

1. J. L. O'Brien, Optical quantum computing. *Science* **318**, 1567-1570 (2007).
2. A. Aspuru-Guzik, P. Walther, Photonic quantum simulators. *Nature Phys.* **8**, 285-291 (2012).
3. E. Knill, R. Laflamme, G. J. Milburn, A scheme for efficient quantum computation with linear optics. *Nature* **409**, 46-52 (2001).
4. D. E. Browne, T. Rudolph, Resource-efficient linear optical quantum computation. *Phys. Rev. Lett.* **95**, 010501 (2005).
5. M. Varnava, D. E. Browne, T. Rudolph, How good must single photon sources and detectors be for efficient linear optical quantum computation? *Phys. Rev. Lett.* **100**, 060502 (2008).
6. T. Jennewein, M. Barbieri, A. G. White, Single photon device requirements for operating linear optics quantum computing outside the post-selection basis. *J. of Mod. Opt.* **58**, 276-287 (2011).
7. M. D. Eisaman, J. Fan, A. Migdall, S. V. Polyakov, Single-photon sources and detectors. *Review of Scientific Instruments* **82**, 071101 (2011).

8. P. Michler et al., A quantum dot single-photon turnstile device. *Science* **290**, 2282-2285 (2000).

9. A. J. Shields, Semiconductor quantum light sources. *Nature Photon.* **1**, 215-223 (2007).

10. S. V. Polyakov *et al.*, Coalesence of single photons emitted by disparate single photon sources: the example of InAs quantum dots and parametric down conversion sources. *Phys. Rev. Lett.* **107**, 157402 (2011).

11. N. Mizuochi et al., Electrically driven single-photon source at room temperature in diamond. *Nature Photon.* **6**, 299–303 (2012).

12. Photon emission from single emitters is on-demand, but inefficient photon collection, filtering and detection introduce random losses of photons and thus result in non-deterministic photon output. As this is a common challenge for most photon sources and is another key area under development, we ignore this effect when we talk about single photon probabilities throughout the paper.

13. K. Konthasinghe et al., Field-field and photon-photon correlations of light scattered by two remote two-level InAs quantum dots on the same substrate. *Phys. Rev. Lett.* **109**, 267402 (2012).

14. A. Sipahigil et al., Indistinguishable photons from separated silicon-vacancy centers in diamond. *Phys. Rev. Lett.* **113**, 113602 (2014).

15. A. Politi, M. J. Cryan, J. G. Rarity, S. Yu, J. L. O'Brien, Silica-on-silicon waveguide quantum circuits. Science **320**, 646-649 (2008).

16. J. B. Spring et al., Boson sampling on a photonic chip. *Science* **339**, 798-801 (2013).

17. B. A. Bell *et al.*, Experimental demonstration of a graph state quantum error-correction code. *Nat. Commun.* **5**, 3658 (2014).

18. X.-C. Yao *et al.*, Observation of eight-photon entanglement. *Nature Photon.* **6**, 225-228 (2012).

19. A. L. Migdall, D. Branning, S. Castelletto, Tailoring single-photon and multiphoton probabilities of a single photon on-demand source. *Phys. Rev. A* **66**, 053805 (2002).

20. T. Pittman, B. Jacobs, J. Franson, Single photons on pseudodemand from stored parametric down-conversion. *Phys. Rev. A* **66**, 042303 (2002).

21. E. Jeffrey, N. A. Peters, P. G. Kwiat, Towards a periodic deterministic source of arbitrary single-photons. *New J. Phys.* **6**, 100 (2004).

22. J. Mower, D. Englund, Efficient generation of single and entangled photons on a silicon photonic integrated chip. *Phys. Rev. A* **84**, 052326 (2011).

23. J. H. Shapiro, F. N. C. Wong, On-demand single-photon generation using a modular array of parametric downconverters with electro-optic polarization controls. *Opt. Lett.* **32**, 2698-2700 (2007).

24. X.-S. Ma, S. Zotter, J. Kofler, T. Jennewein, A. Zeilinger, Experimental generation of single photons via active multiplexing. *Phys. Rev. A* **83**, 043814 (2011).

25. M. J. Collins *et al*., Integrated spatial multiplexing of heralded single-photon sources. *Nat. Commun.* **4**, 2582 (2013).

26. D. Bonneau, G. J. Mendoza, J. L. O'Brien, M. G. Thompson, Effect of loss on multiplexed single-photon sources. *New J. Phys.* **17**, 043057 (2015).

27. See supporting material on Science Online.

28. X. Zhang *et al.*, Enhancing the heralded single-photon rate from a silicon nanowire by time and wavelength division multiplexing pump pulses. *Opt. Lett.* **40**, 2489-2492 (2015).

29. H. Takesue, K. Shimizu, Effects of multiple pairs on visibility measurements of entangled photons generated by spontaneous parametric processes. *Opt. Commun.* **283**, 276-287 (2010).

30. C. K. Hong, Z. Y. Ou, L. Mandel, Measurement of subpicosecond time intervals between two photons by interference. *Phys. Rev. Lett.* **59**, 2044-2046 (1987).

31. K. Harada *et al.*, Indistinguishable photon pair generation using two independent silicon wire waveguides. *New J. Phys.* **13**, 065005 (2011).

32. A. R. McMillan et al., Two-photon interference between disparate sources for quantum networking. *Sci. Rep.* **3**, 2032 (2013).

**Acknowledgments:** We thank Bryn Bell and Jiakun He for helpful discussion. We acknowledge funding support from the Centre of Excellence (CUDOS, CE110001018), Laureate Fellowship (FL120100029), Discovery Early Career Researcher Award (DE120100226), Future Fellowship (FT110100853), Discovery Project (DP130100086), and Linkage Project (LP130101034) programs of the Australian Research Council (ARC).

## Materials and Methods

Full setup and explanation of coincidence measurements

The full setup of our experiments is illustrated in Fig. S1. The top half of the diagram, excluding the 50:50 coupler before the two avalanche photo diode single photon detectors (ID210, Id-Quantique), is for the CAR measurements of the source with four-fold temporal-mode multiplexing (MUX). When doing the measurements for the source without MUX, we connected only one of the four channels of the 1-to-4 fiber couplers (FCs), and removed the switching network. This arrangement allowed the least change to the experimental conditions for the MUX and NO MUX measurements, and thus guaranteed a fair comparison between them. Because the losses of the four channels of the FCs were slightly different, we used one polarization controller (PC) in each channel and an inline polarizer immediately after the FCs to ensure the pulses in the four temporal modes after the inline polarizer had the same intensity. A fast optical sampling oscilloscope was used to monitor the pulse intensities. The CAR measurements for pumping at the 10 MHz clock of only $t_1$, $t_2$, $t_3$ or $t_4$ (Fig. 1) indicate that all four NO MUX sources have the same performance, as shown by diamonds in Fig. 3A.

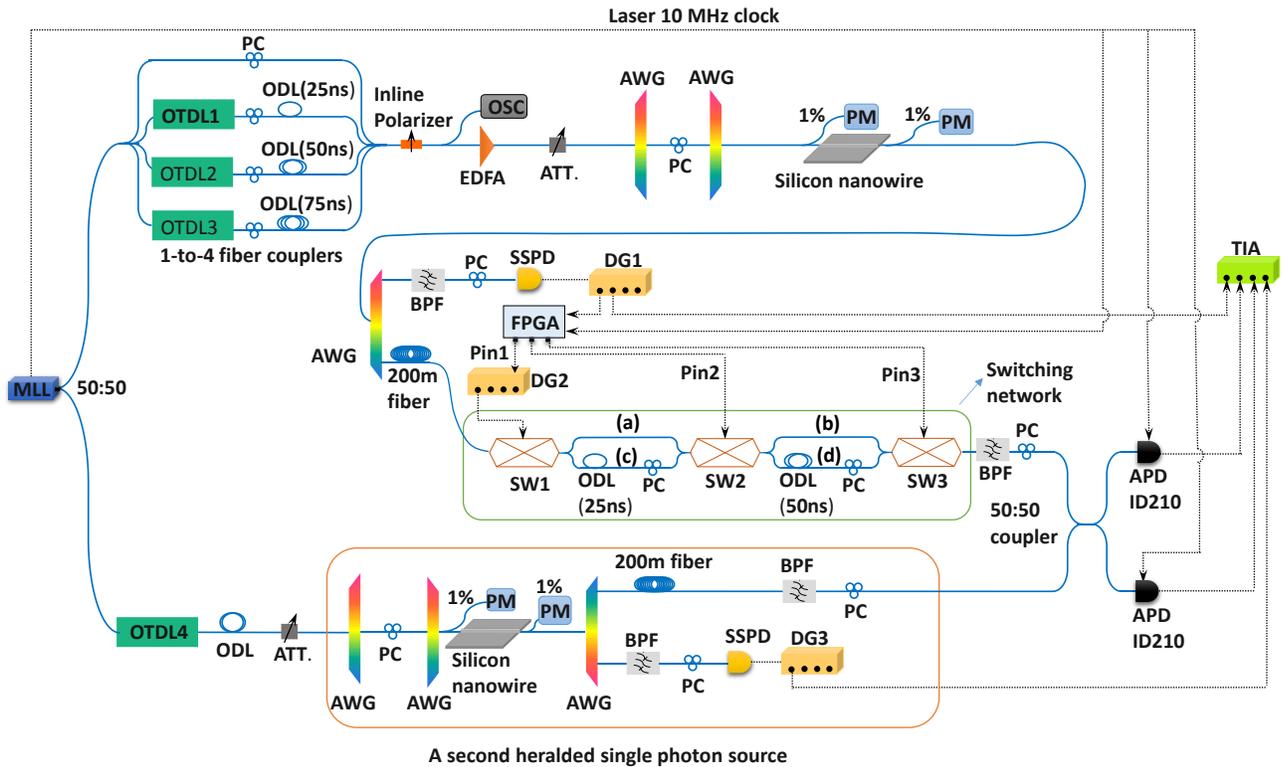

**Fig. S1.**
The full setup of the experiments. Solid and dashed lines represent optical fibers and electronic cables, respectively. MLL: mode-locked laser, OTDL: optical tunable delay line, ODL: optical delay line, PC: polarization controller, ATT: attenuator, OSC: oscilloscope, AWG: arrayed waveguide grating, PM: power meter, SSPD: superconducting single photon detector, DG: delay generator, FPGA: field-programmable gate array, SW: switch, BPF: bandpass filter, APD: avalanche photon diode, TIA: time interval analyzer.

The bottom half of Fig. S1 is a second heralded single photon source based on a silicon nanowire (*28*) with the same specification to that used in the MUX experiments. This source was pumped at the laser's 10 MHz clock, and provided a reference for the indistinguishability test of the multiplexed photons via the four-fold HOM quantum interference (*31*).

When we performed the CAR measurements for the NO MUX and MUX photon sources, we took the 50:50 coupler out of the setup and connected the heralded photon output directly to one ID210 detector. In both cases, the heralded photon events detected by an ID210 triggered by the 10 MHz laser clock were used as the 'start', and the heralding photon events detected by a SSPD were used as the 'stop' for the TIA to construct the histograms: coincidences vs time delay between 'start' and 'stop'. This 'start' and 'stop' arrangement is different from the standard way of using heralding signals as the 'start' and heralded as the 'stop' just for our experimental convenience to adjust the delays, and it does not change the physics. A suitable electronic delay was applied to the heralding detection signals so that the delay in the histogram was always within the 0–250 ns time window.

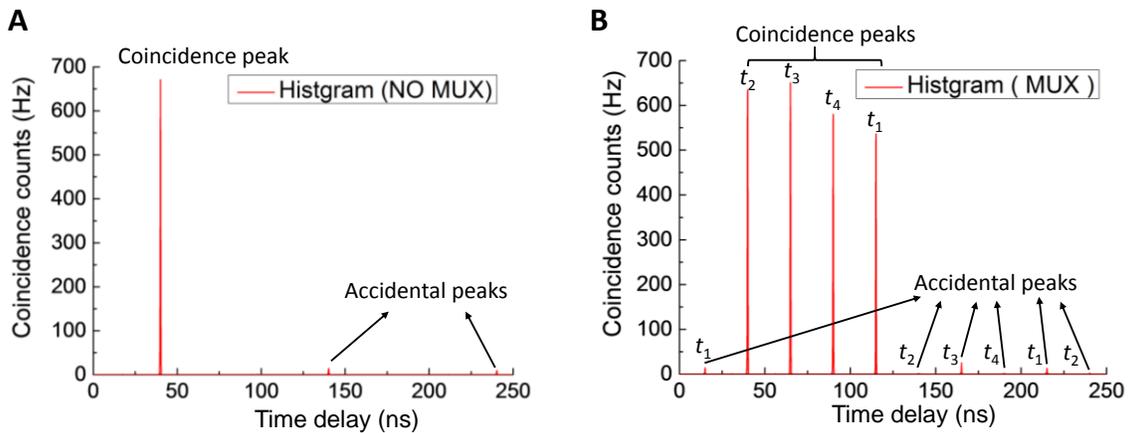

**Fig. S2**
Typical histograms for CAR measurements. (A) NO MUX: one coincidence peak. (B) MUX: four coincidence peaks represent photons from four time bins shown in Fig. 1 and two accidental peaks (two nearly disappear because of switching).

Figure S2 shows typical histograms for the NO MUX and MUX sources in the CAR measurements (*28*). For the NO MUX source, we sum up all counts in the coincidence peak and the accidental peak closest to the coincidence peak as the coincidences and accidentals, respectively. The time interval between the peaks is the pump laser period. The ratio between them gives the CAR. For the MUX source, depending on in which time bin the heralded photons are generated, they experience different optical delays to be multiplexed to the 10 MHz clock; however the heralding photons are still on the 40 MHz clock, so there are four coincidence peaks (the signature of successful multiplexing). These peaks look slightly different from one another because photons generated in different time bins experience slightly different losses when propagating through different optical paths. Each coincidence peak should have their corresponding accidental peak. Two accidental peaks nearly disappear because the photons that are generated in time bins $t_2$ and $t_4$ require logic '1' be applied to switch1 (Fig. 2). This switching operation on switch1 only allows heralded photons to arrive at the detector when there are heralding events, and therefore switches away all other events that may give rise to coincidences in the accidental peak. To calculate the CAR for the MUX source, we sum up the counts in all four coincidence peaks as the coincidences; and multiply the overall counts in the two visible accidental peaks by two as the accidentals. This process is considered to be fair because: first, the sources pumped by each individual temporal mode have been tested to have the same performance; second, corresponding to the loss differences between the different paths, the accidentals for $t_3$ should be higher than those for $t_2$, while the accidentals for $t_1$ should be lower than those for $t_4$, and the total accidentals for $t_3$ and $t_1$ can approximately represent those for $t_2$ and $t_4$.

Clock, optical delay, and switching network loss management

To make this temporal multiplexing experiment successful, we had to synchronize: (I) the 40 MHz clock of the FPGA and the heralding event so that the FPGA could find out which time bin the heralding photons were generated; (II) the switching electronic signals from the FPGA and the heralded photons arriving at the switches so that proper delays could be applied to the photons; and (III) the relative optical delays in the four channels of the FCs and in the switching network.

For synchronization (I), we used a digital delay generator (DG1 in Fig. S1) between the SSPD heralding detection signal output and the FPGA. By doing an ADD operation between the heralding detection signals and the FPGA clock, and comparing the FPGA output count rate with that on the SSPD software, we could find the correct delay using DG1. Once this was done, we could proceed to synchronization (II).

To simplify synchronization (II), we configured the FPGA such that according to a heralding event, the three binary-digit output was latched until the next heralding event occurred. For example, if a photon pair was generated in time bin $t_1$, '000' was applied to the switches until the next pair was generated. If the next pair was generated in time bin $t_4$, the output of the FPGA would change to '110' (Fig. 2). In this way we only need to synchronize the switching signals that control switch1 with the incoming photons, because once photons pass through switch1, they are routed directly through switch2 and switch3 that are always under the correct logic control.

Because the relative delay between heralding and heralded events was the same no matter in which time bin the photon pairs were generated, we only need to do synchronization for one pump temporal mode. As the initial input status to the switches was '0', we had to choose the pump on the clock of either $t_2$ or $t_4$ for synchronization because only these two pump modes required a logic '1' input to switch1 and were possible for delay alignment. In the setup shown in Fig. S1, we only connected the channel of FCs with 25 ns optical delays (i.e. $t_2$) and applied a constant output '101' from the FPGA to the switches to take a coincidence measurement as the reference. Then we added a digital delay generator (DG2 in Fig. S1) between the logic output pin1 of the FPGA and switch1. We could continuously tune the delay of the switching signals to switch1 via DG2 and measure the coincidences. By trial and error, the correct delay was found when the measured coincidences matched with the reference measurement. This delay was then double confirmed by coincidence measurements for the other three temporal pump modes.

For synchronization (III), we manually cut optical fibers with the lengths providing delays of approximately 25 ns and 50 ns and spliced the fibers in paths (c) and (d) of the switching network; and then used off-the-shelf fiber-integrated optical tunable delay lines (OTDL) with a tuning step of 1 ps in three channels of the FCs for multiplexing the pump to match the delays in the switching network. Compared with directly using OTDL in the switching network, this arrangement minimized the losses of the multiplexed heralded photons.

Polarization management

Due to the use of non-polarization maintaining components in the switching network, photons with the same polarization at the input generally will have different polarization at the output if they go through different optical paths. Depending on in which time bin the heralded photons are generated, they go through the combination of optical paths, $t_1$: (a)+(b), $t_2$: (c)+(b), $t_3$: (a)+(d), and $t_4$: (c)+(d). We found that adding one PC in each of (c) and (d) was sufficient to maintain the polarization of the photons. For example, the photons from $t_1$ and $t_2$ share the path (b), the PC in (c) can always adjust the polarization of photons to be the same to those in (a). Then (c) becomes equal to (a) in terms of polarization. The same rule applies to (d) and (b) and thus photons from all time bins will have the same polarization at the output of the switching network.

Inferring heralded single photon output probability from coincidences

Because we did not use photon number resolving detectors, we cannot measure the heralded single photon output probability directly, but we can infer it from the coincidences measured by threshold detectors. As shown in Fig. S1, both the generated photons and the pump were filtered using AWGs with the same channel bandwidth of 50 GHz. This arrangement makes it appropriate to use the thermal distribution to describe the photon generation statistics (*29*). Using the measured coincidences, we calculate the mean number of generated pairs per 100 ns clock period $\mu$, by taking into account the total losses in the heralding and heralded photon arms. Using $P_1=\mu/(1+\mu)^2$ we infer the heralded single photon generation probability per 100 ns clock period. Using $P_1\eta$ with $\eta$ being the overall photon pair collection efficiency, we calculate the heralded single photon output probability for the plots in Fig. 3B.

Four-fold HOM interference measurements

The delay management procedure described in a previous section "Clock, optical delay, and switching network loss management" aligns the photon arrival time to an accuracy of 1 ns, which is determined by the time resolution of the coincidence measurement system. The four-fold HOM interference requires the delay alignment at the accuracy of photons' coherence time which is of the order of the pump pulse width 10 ps. The measurements were taken in the following steps. First, we connected only the channel of the 1-to-4 FCs without optical delay lines as the pump of the MUX source so that photons were always generated in time bin $t_1$. By varying the delay of OTDL4 in the second source, we adjusted the fine delay via two-fold HOM interference and fixed the delay on OTDL4. Second, we connected in turn only the channel of the 1-to-4 FCs with OTDL1, OTDL2, or OTDL3, to determine and fix the fine delays for them via two-fold measurements. Third, we connected all four channels of the 1-to-4 FCs to obtain the fully multiplexed source and took the two-fold and four-fold HOM interference measurements by varying the fine delay on OTDL4.

As the pump powers for the four-fold HOM interference measurement were relatively high, we attribute the 69% visibility predominantly to multi-pair noise. To obtain this noise information, we performed four-fold coincidence measurements at the same fine delays set in the raw measurements by disconnecting input to the 50:50 coupler from the MUX and second photon sources, respectively (*32*). Because subtracting these four-fold counts from the raw four-fold counts would subtract the noise due to detector dark count twice, we also measured the four-fold coincidences by disconnecting both sources from the 50:50 coupler, and added these dark count induced four-fold counts back to get the net four-fold coincidences. To understand this process, we give an example: we obtained a raw four-fold coincidence count $C_{raw}$ at a particular delay $\delta t$.

We measured the multi-pair noise contribution to $C_{\text{raw}}$ from two sources at delay $\delta t$ to be $C_{n1}$ and $C_{n2}$, and the detector dark count contribution to $C_{\text{raw}}$ at delay $\delta t$ to be $C_d$. The corrected net four-fold coincidence is then $C_{\text{raw}} - C_{n1} - C_{n2} + C_d$. We did this correction for all delays shown in Fig. 4A and obtained the data for Fig. 4B. This multi-pair noise subtraction yields a 91% visibility.